\documentclass[aps,prl,reprint,preprintnumbers,superscriptaddress,amsmath,amssymb,bibnotes,longbibliography]{revtex4-1}
\usepackage{graphicx}
\usepackage{dcolumn}
\usepackage{epstopdf}
\usepackage{braket}
\usepackage{bm,multirow,threeparttable}
\usepackage{flushend}
\usepackage[pdfstartview=FitH, CJKbookmarks=true, bookmarksnumbered=true, bookmarksopen=true, colorlinks, pdfborder=001, linkcolor=blue, anchorcolor=blue, citecolor=blue,urlcolor=blue]{hyperref}
\newcommand{\red}[1]{\textcolor{black}{#1}}

\begin{document}
\title{Structural and physical properties of the heavy fermion metal Ce$_2$NiAl$_6$Si$_5$}
\author{Jiawen Zhang}
\affiliation  {Center for Correlated Matter and School of Physics, Zhejiang University, Hangzhou 310058, People’s Republic of China}
\author{Jinyu Wu}
\affiliation  {Center for Correlated Matter and School of Physics, Zhejiang University, Hangzhou 310058, People’s Republic of China}
\author{Ye Chen}
\affiliation  {Center for Correlated Matter and School of Physics, Zhejiang University, Hangzhou 310058, People’s Republic of China}
\author{Rui Li}
\affiliation  {Center for Correlated Matter and School of Physics, Zhejiang University, Hangzhou 310058, People’s Republic of China}
\author{Michael Smidman}
\affiliation  {Center for Correlated Matter and School of Physics, Zhejiang University, Hangzhou 310058, People’s Republic of China}
\author{Yu Liu}
\email[Corresponding author: ]{liuyu002@mail.ustc.edu.cn}
\affiliation  {Center for Correlated Matter and School of Physics, Zhejiang University, Hangzhou 310058, People’s Republic of China}
\author{Yu Song}
\email[Corresponding author: ]{yusong\_phys@zju.edu.cn}
\affiliation  {Center for Correlated Matter and School of Physics, Zhejiang University, Hangzhou 310058, People’s Republic of China}
\author{Huiqiu Yuan}
\email[Corresponding author: ]{hqyuan@zju.edu.cn}
\affiliation  {Center for Correlated Matter and School of Physics, Zhejiang University, Hangzhou 310058, People’s Republic of China}
\affiliation  {State Key Laboratory of Silicon and Advanced Semiconductor Materials, Zhejiang University, Hangzhou 310058, People’s Republic of China}
\affiliation{Collaborative Innovation Center of Advanced Microstructures, Nanjing 210093, People’s Republic of China}

\date{\today}

\begin{abstract}
Strongly correlated electrons at the verge of quantum criticality give rise to unconventional phases of matter and behaviors, with the discovery of new quantum critical materials driving synergistic advances in both experiments and theory. 
In this work, we report the structural and physical properties of a new quaternary Ce-based heavy fermion compound, Ce$_2$NiAl$_6$Si$_5$, synthesized using the self-flux method. This compound forms a layered tetragonal structure (space group $P4/nmm$), with square nets of Ce atoms separated by Si-Al or Ni-Si-Ge layers. Specific heat measurements show a low-temperature Sommerfeld coefficient of 1.4 J/mol-Ce K$^2$, with a reduced entropy indicative of significant Kondo interactions. Below 0.6~K, an upturn in resistivity and a deviation in magnetic susceptibility suggest the appearance of magnetic ordering or the development of dynamic magnetic correlations, which is further supported by a bulge in specific heat around 0.4~K.
These results suggest that Ce$_2$NiAl$_6$Si$_5$ is a layered heavy fermion metal, naturally located in proximity to a spin-density-wave quantum critical point.

\begin{description}
\item[PACS number(s)]
\end{description}
\end{abstract}

\maketitle

\section{Introduction}

Strong electronic correlations lead to an enhanced electron effective mass and narrowed electronic band widths, as exemplified in heavy fermion metals, which can have electrons (fermions) with effective mass over 1000 times the free electron mass, resulting from the hybridization between localized $f$-electrons and light conduction electrons \cite{RevModPhys.56.755}. The relatively small energy scales in heavy fermion metals allow for effective tuning of their ground states via non-thermal parameters such as pressure, magnetic fields, chemical doping and dimensionality \cite{Weng_2016,Chen_2016,Smidman_2019,Xin2018}. Magnetically ordered phases can be continuously suppressed through these non-thermal parameters, leading to quantum critical points (QCPs) \cite{Weng_2016}, around which unconventional superconductivity and non-Fermi-liquid (NFL) behaviors are observed \cite{WHITE2015246}. Such QCPs have been extensively studied in antiferromagnetic (AFM) heavy fermion metals (such as CeRhIn$_5$ \cite{ParkT2006,Kne2006,Jiao2015}, CePd$_2$Si$_2$ \cite{MathurND1998}, and CeCu$_{6-x}$Au$_x$ \cite{Sto2011}), and recently in ferromagnetic (FM) CeRh$_6$Ge$_4$ \cite{MatsuokaE2015,ShenB2020,Kot2019}. The high tunability and the plethora of ordered phases in heavy fermion metals make them ideal for studying quantum phase transitions and unconventional states of matter that emerge around QCPs \cite{Stew2001,L2007,Pfl2009,WHITE2015246,Si2010,Weng_2016}, and motivate the search for new materials that are close to QCPs. Compounds with quasi-two-dimensional (2D) square lattices are particularly interesting, as they may host physics similar to high-$T_{\rm c}$ superconducting cuprates \cite{Da1994,LeggettAJ2006} and iron pnictides/chalcogenides \cite{FernandesRM2022,SiQ2016,BL2022}. 

Recently, the family of Ce$_2M$Al$_7$Ge$_4$ ($M$ = Co, Ir, Ni, Pd) compounds is found to exhibit intriguing physical properties \cite{PhysRevB.93.205141}. 
These compounds crystallize in a noncentrosymmetric tetragonal structure (space group $P\bar{4}2_1m$), with quasi-2D Ce layers separated by Ge-Al and $M$-Al-Ge blocks \cite{PhysRevB.93.205141}. Ce$_2$PdAl$_7$Ge$_4$ is suggested to be at a QCP with NFL behaviors extending down to 0.4~K, while the other variants with $M$~=~Co, Ir, and Ni order magnetically at 1.8~K, 1.6~K, and 0.8~K, respectively \cite{PhysRevB.93.205141}. In this work, we successfully synthesized a new heavy fermion metal Ce$_2$NiAl$_6$Si$_5$, discovered in the course of exploring chemical pressure effects in Ce$_2$NiAl$_7$Ge$_4$, via partially replacing Ge with Si.
Unlike the Ce$_2$$M$Al$_7$Ge$_4$ compounds, Ce$_2$NiAl$_6$Si$_5$ crystallizes in a centrosymmetric tetragonal structure (space group $P4/nmm$). 
Around 0.6~K, an upturn in resistivity and a deviation from Curie-Weiss behavior in magnetic susceptibility suggest the onset of \red{AFM} order or dynamic \red{AFM} correlations, also corroborated by a bulge in specific heat that occurs at al slightly lower temperature. A large Sommerfeld coefficient of 1.4~J/mol-Ce~K$^2$ at 0.1~K and the reduction of entropy evidence significant Kondo interactions in Ce$_2$NiAl$_6$Si$_5$. These results indicate that Ce$_2$NiAl$_6$Si$_5$ is a square-lattice heavy fermion metal located close to a QCP, and motivates further pressure- or field-tuning studies that directly access the QCP. 

\section{Experimental methods}
Single crystals of Ce$_2$NiAl$_6$Si$_5$ were synthesized by the Al/Si self-flux method. Ce rods (Alfa Aesar 99.9\%), Ni powder (PrMat 99.99\%), Si ingots (PrMat 99.9999\%), and Al granules (PrMat 99.9999\%), were placed in an alumina crucible in a molar ratio of Ce~:~Ni~:~Al~:~Si $ = 1 : 1 : 35 : 8$ and sealed in an evacuated quartz ampoule. The ampoule was heated up to 1100~$^\circ$C, held at this temperature for 24~h and then slowly cooled to 700~$^\circ$C. Finally, shiny platelet single crystals of Ce$_2$NiAl$_6$Si$_5$ were obtained after centrifuging to remove the excess flux. 

The chemical composition was determined using energy-dispersive spectroscopy with a Hitachi SU8010 scanning electron microscope, which gives an atomic ratio of Ce~:~Ni~:~Al~:~Si~=$14.82 : 7.55 : 40.29 : 37.34$ ($2: 1.02 : 5.44 : 5.04$). The crystal structure was characterized using single-crystal X-ray diffraction (XRD) with a Bruker D8 Venture diffractometer with Mo $K_\alpha$ radiation at room temperature. The electrical resistivity and heat capacity were measured in a Quantum Design Physical Property Measurement System (PPMS) equipped with a dilution refrigerator. Magnetization measurements were performed using a Quantum Design Magnetic Property Measurement System (MPMS) equipped with a $^3$He refrigerator.

\section{Results and discussion}
Refinement of single crystal X-ray diffraction results indicates that Ce$_2$NiAl$_6$Si$_5$ crystallizes in a defect variant of the Sm$_2$NiGa$_{12}$ structure \cite{ChenXZ2000,ChoJY2008,MACALUSO20053547}, with a centrosymmetric space group $P4/nmm$ (No.129). Although the noncentrosymmetric space group $P\bar{4}2_1m$ yields a better refinement in the Ce$_2$$M$Al$_7$Ge$_4$ family \cite{PhysRevB.93.205141}, the higher-symmetry $P4/nmm$ space group yields a better refinement than the $P\bar{4}2_1m$  space group ($w R_2 = 0.0736$) for Ce$_2$NiAl$_6$Si$_5$.

The refined crystal structure for Ce$_2$NiAl$_6$Si$_5$ is presented in Table~\ref{table1}. In this crystal structure, there is a single Ce site and a single Ni site, at the 4$f$ and 2$b$ Wyckoff positions, respectively. There are two crystallographically inequivalent Al sites and three inequivalent Si sites. 
As in typical intermetallic compounds containing Al and Si, it is difficult to distinguish these two elements using X-ray scattering \cite{ChenXZ1998}. We find that the model in Table~\ref{table1} gives the lowest $R$-value in the refinement, although a similar $R$-value ($w R_2 = 0.0401$) is possible if Al($3$) is assigned as Si, but is ruled out because it leads to a chemical composition inconsistent with EDS measurements.
While our X-ray diffraction measurements favor the crystal structure presented in Table~\ref{table1}, further neutron diffraction measurements are desirable to better distinguish between Al and Si.

\begin{table}[!ht]
	\renewcommand\arraystretch{1.4}
	\caption{Crystal structure of Ce$_2$NiAl$_6$Si$_5$ determined from single crystal X-ray diffraction measurements at 300~K with Mo $K_\alpha$ radiation (0.71073~\AA): $a = 5.86260(10)$~\AA, $c = 14.7892(5) $~\AA, $V = 508.31(2) $~\AA$^3$, space group $P4/nmm$ (No.129). Goodness of fit is $1.299$, $R_1 = 0.0209$, $w R_2 = 0.0403$ (all data) . The refined structural parameters and isotropic displacement parameters $U_{\rm eq}$ of Ce$_2$NiAl$_6$Si$_5$ are shown below, where $U_{\rm eq}$ is taken as $1/3$ of the trace of the orthogonalized $U_{ij}$ tensor.}
	\begin{tabular}{lcccccc}
		\hline\hline
		Atoms & Wyck. & ~~$x$~~ & ~~$y$~~ & ~~$z$~~ & Occ. & $U_{\rm eq}$(\AA$^2$) 
		\\ \hline
		Ce(1) & $4f$ & 0.2500 & 0.7500 & 0.2357(1) & 1 & 0.0075(2) 
		\\
		Ni(1) & $2b$ & 0.7500 & 0.2500 & 0.5000 & 1 & 0.0071(2) 
		\\
		Al(1) & $8j$ & 0.4958(1) & 0.4958(1) & 0.4118(1) & 1 & 0.0081(2) 
		\\
		Al(2) & $2c$ & 0.7500 & 0.7500 & 0.1361(1) & 1 & 0.0091(4) 
		\\
		Al(3) & $2c$ & 0.2500 & 0.2500 & 0.2974(1) & 1 & 0.0043(4) 
		\\
		Si(1) & $2c$ & 0.7500 & 0.7500 & 0.3057(1) & 1 & 0.0080(4)
		\\
		Si(2) & $8j$ & 0.4572(1) & 0.4572(1) & 0.0771(1) & 1 & 0.0140(2) 
		\\ \hline\hline
	\end{tabular}
	\label{table1}
\end{table}

\begin{figure*}[htbp]
	\begin{center}
		\includegraphics[width=\textwidth]{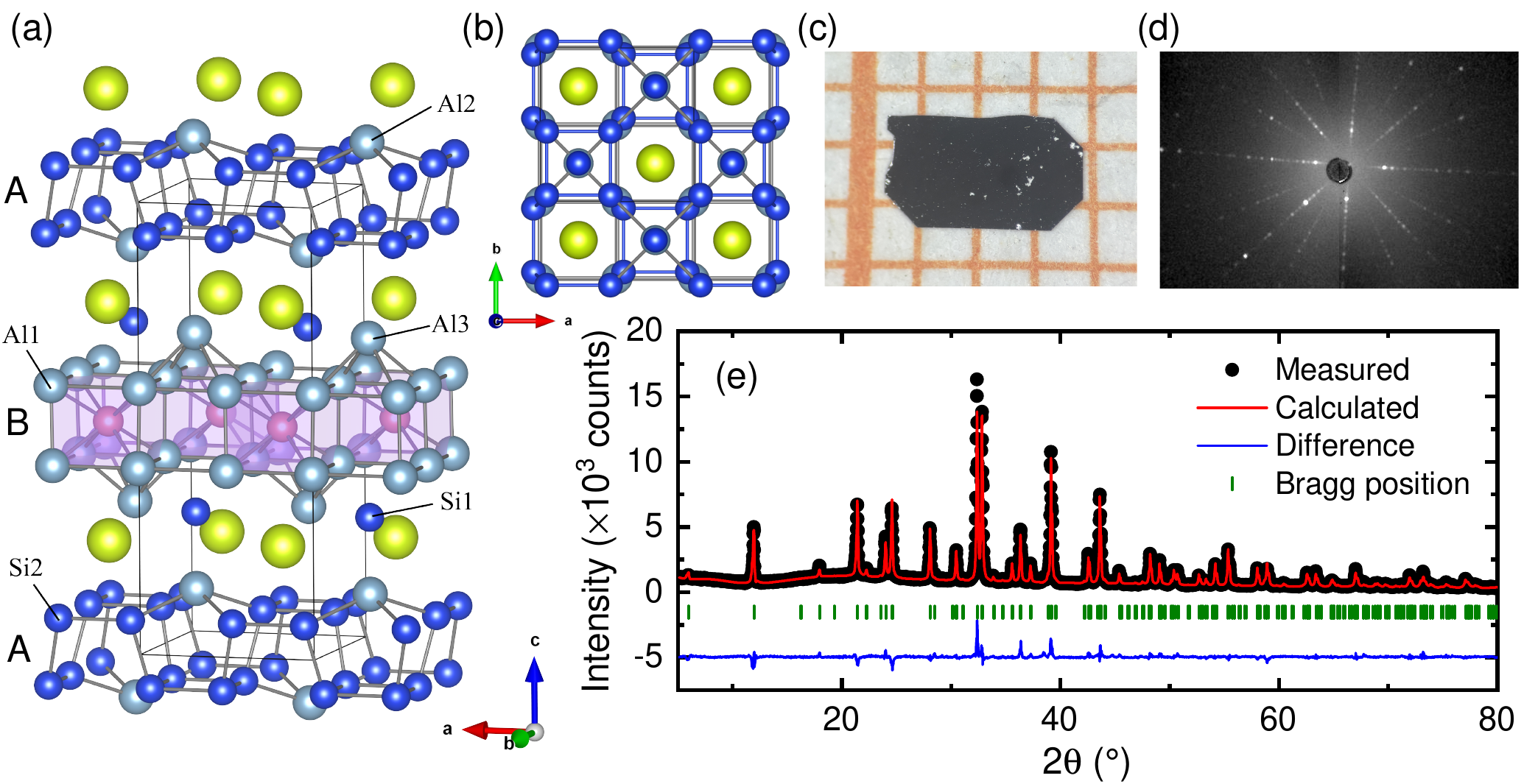}
	\end{center}
	\caption{(Color online) (a)-(b) Crystal structure of Ce$_2$NiAl$_6$Si$_5$, where green, orange, gray and blue spheres represent Ce, Ni, Al and Si atoms, respectively. (c) Image of a Ce$_2$NiAl$_5$Si$_6$ crystal on a 1~mm~$\times$~1~mm grid. (d) Laue pattern of Ce$_2$NiAl$_6$Si$_5$. (e) Powder XRD data of Ce$_2$NiAl$_6$Si$_5$ at room temperature, with the corresponding Rietveld refinement shown as a solid line.}
	\label{figure1}
\end{figure*}

The crystal structure of Ce$_2$NiAl$_6$Si$_5$ is shown in Figs.~\ref{figure1}(a) and (b), with planar square nets of Ce atoms separated by alternating Al-Si (A-type) and Ni-Al-Si (B-type) layers along the $c$-axis. \red{The intra-layer distance between Ce atoms is 4.15~{\AA}, whereas the inter-layer distances between Ce atoms are 6.97~{\AA} (separated by the A-type layer) and 7.82~{\AA} (separated by the B-type layer), suggesting Ce$_2$NiAl$_6$Si$_5$ to be a quasi-2D square lattice compound.}
The A-type layer is formed by Al(2) and Si(2) atoms, which is similar to the Al-Si layer in Sm$_2$Ni(Ni$_x$Si$_{1-x}$)Al$_4$Si$_6$ \cite{ChenXZ1998}. Compared to the A-layer (formed by Al and Ge) in Ce$_2$NiAl$_7$Ge$_4$ that leads to a noncentrosymmetric structure, the A-layer in Ce$_2$NiAl$_6$Si$_5$ has a different Al-Si pattern and results in an overall centrosymmetric structure. The B-type layer consists of five atomic layers stacked in the sequence Al/Si-Al-Ni-Al-Al/Si. The Ni atoms are surrounded by Al atoms, forming NiAl$_8$ distorted cubes. These NiAl$_8$ cubes are similar to the NiGa$_8$ cubes in Sm$_2$NiGa$_{12}$ \cite{ChenXZ2000}, NiAl$_8$ cubes in Ce$_2$NiAl$_7$Ge$_4$ \cite{PhysRevB.93.205141}, and PtAl$_8$ cubes in CePtAl$_4$Si$_2$ \cite{Ghimire_2015}. Above and below the NiAl$_8$ cubes are Al(3) and Si(1) atoms that have very close $z$ coordinates. Overall, the crystal structure of Ce$_2$NiAl$_6$Si$_5$ is similar to that of Ce$_2$NiAl$_7$Ge$_4$, with the main difference being the Al(3) and Si(1) sites in Ce$_2$NiAl$_6$Si$_5$ are both occupied by Ge atoms in Ce$_2$NiAl$_7$Ge$_4$, which also changes the atomic ratio of elements from 2~:~1~:~6~:~5 to 2~:~1~:~7~:~4. 


The crystal structure of Ce$_2$NiAl$_6$Si$_5$ obtained from single crystal X-ray diffraction fully accounts for its powder X-ray diffraction pattern, as shown in Fig. \ref{figure1}(e), and further shows that the synthesized samples are single-phase with minimal excess flux.
We note that while Al(3) and Si(1) are ordered in our refined structure, we do not rule out mixing between the two sites, due to the similarity between Al and Si as observed by XRD. It is also possible that there might be Al deficiencies in our sample, as energy-dispersive spectroscopy measurements suggest a value of $\sim5.4$, deviating from the ideal value of 6.

\begin{figure}[htbp]
	\begin{center}
		\includegraphics[width=\columnwidth]{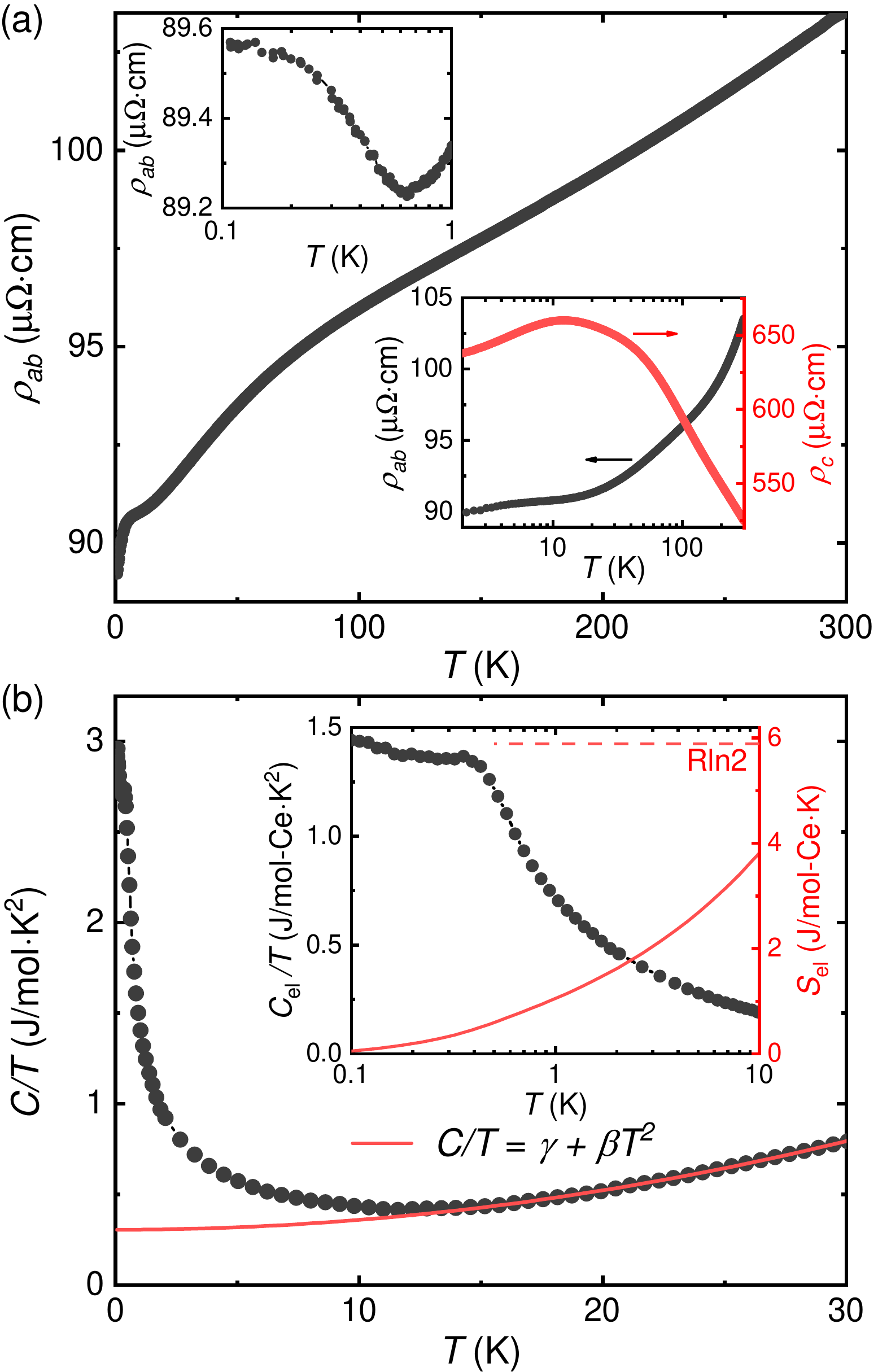}
	\end{center}
	\caption{(Color online) (a) Temperature dependence of the resistivity $\rho_{ab}(T)$ of Ce$_2$NiAl$_6$Si$_5$. \red{The upper inset zooms into $\rho_{ab}$ below 1 K. The lower inset compares $\rho_{ab}$ (left axis) and $\rho_{c}$ (right axis).} (b) Temperature dependence of the specific heat $C(T)/T$ of Ce$_2$NiAl$_6$Si$_5$. The inset shows the electron specific heat $C_{\rm el}(T)/T$ (left axis) and entropy $S_{\rm el}$ (right axis).}
	\label{figure2}
\end{figure}

Fig.~\ref{figure2}(a) shows temperature dependence of the electrical resistivity $\rho_{ab}$ of Ce$_2$NiAl$_6$Si$_5$, with a broad hump around 100~K. At low temperatures, $\rho_{ab}$ drops sharply below 5~K and exhibits an upturn at around 0.6~K, as shown in the \red{upper inset} of Fig.~\ref{figure2}(a). This upturn in $\rho(T)$ suggests the opening of an electronic gap, as seen in A-type CeCu$_2$Si$_2$ \cite{Ge1998}, YbBiPt \cite{Fisk1991}, and CeRu$_2$Si$_2$ \cite{MiyakoY1997}, and may result from spin-density-wave magnetic ordering in Ce$_2$NiAl$_6$Si$_5$ below 0.6~K. \red{As shown in the lower inset of Fig.~\ref{figure2}(a), the inter-layer resistivity $\rho_{c}$ is significantly larger than $\rho_{ab}$. Upon cooling, $\rho_c$ first increases and then decreases below 11 K, in stark contrast to $\rho_{ab}$. The strong anisotropy between of $\rho_{ab}$ and $\rho_c$ is characteristic of quasi-2D systems, and is consistent with the layered structure of Ce$_2$NiAl$_6$Si$_5$.}

Temperature dependence of the specific heat $C(T)/T$ of Ce$_2$NiAl$_6$Si$_5$ is plotted in Fig.~\ref{figure2}(b). In the paramagnetic state between 15~K and 30~K, $C(T)/T$ was fitted using $C(T)/T = \gamma + \beta T^2$, yielding a Sommerfeld coefficient $\gamma = 149.5(5)$~mJ/mol-Ce~K$^2$ and $\beta = 0.532(2)$~mJ/mol-Ce~K$^4$. A Debye temperature of $\Theta_{\rm D} = 366.7$~K was calculated using $\Theta_{\rm D} = \sqrt[3]{12\pi^4nR/5\beta}$, where $n \approx 13.4$ is the number of atoms per formula unit, and $R = 8.314$~J/mol~K. Below 12~K, $C(T)/T$ increases with decreasing temperature, and the electronic contribution to the specific heat $C_{\rm el}(T)/T$ was estimated by subtracting the fitted phonon contribution. As shown in the inset of Fig.~\ref{figure2}(b), a significant change in the slope of $C_{\rm el}(T)/T$ occurs around 0.4~K. 
While $C_{\rm el}(T)/T$ does not reduce upon further cooling below 0.4~K, as expected in magnetically well-ordered metals, such a change in slope is consistent with magnetic ordering or the growth of magnetic correlations in heavy fermion systems near QCPs, with similar features seen in YbBiPt \cite{Fisk1991}, CeCu$_{1-x}$Au$_x$ \cite{HVL1996}, and CeRh$_6$Ge$_{4-x}$Si$_x$ \cite{Yongjun2022}.
At the lowest measured temperature, $C_{\rm el}(T)/T$ reaches a large value of $\gamma$~= 1.4~J/mol-Ce~K$^2$, due to contributions from heavy electrons. The electronic entropy $S_{\rm el}$ was obtained by integrating $C_{\rm el}(T)/T$ with respect to $T$ and shown in the inset of Fig.~\ref{figure2}(b). The electronic entropy of Ce$_2$NiAl$_6$Si$_5$ per Ce is only 0.65 of $R$ln2 at 10~K, and recovers to $R$ln2 at around 20~K (not shown). This implies significant Kondo screening of the Ce moments or magnetic correlations between the Ce moments below 10~K.

\begin{figure}[htbp]
	\begin{center}
		\includegraphics[width=\columnwidth]{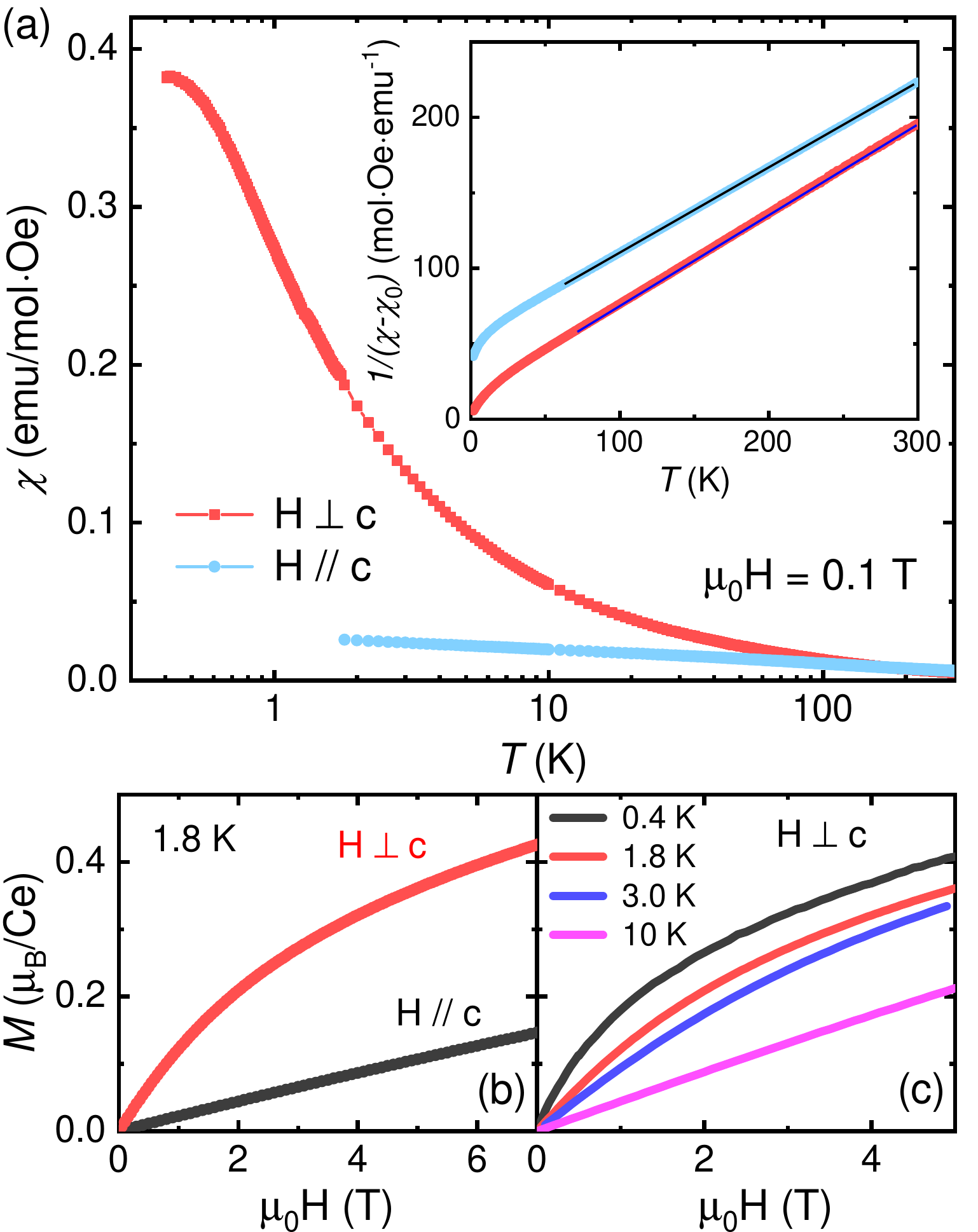}
	\end{center}
	\caption{(Color online) (a) Temperature dependence of the magnetic susceptibility $\chi(T)$ at 0.1~T for $H \parallel c$ and $H \perp c$. The inset shows the inverse magnetic susceptibility $1/(\chi-\chi_0)$ for the two field directions, and the solid lines are Curie-Weiss fits. (b) Field dependence of the magnetization of Ce$_2$NiAl$_6$Si$_5$ at 1.8~K for $H \parallel c$ and $H \perp c$. (c) Isothermal magnetization of Ce$_2$NiAl$_6$Si$_5$ at various temperatures for $H \perp c$.}
	\label{figure3}
\end{figure}

Fig.~\ref{figure3}(a) displays the temperature dependence of the magnetic susceptibilities of Ce$_2$NiAl$_6$Si$_5$, measured at $\mu_0H = 0.1$~T, for $H \parallel c$ and $H \perp c$ [$\chi_\parallel(T)$ and $\chi_\perp(T)$]. Below 200~K, $\chi_\perp(T)$ is larger than $\chi_\parallel(T)$, indicating an easy-plane magnetic anisotropy. As shown in the inset of Fig. \ref{figure3}(a), both $\chi_\parallel(T)$ and $\chi_\perp(T)$ of Ce$_2$NiAl$_6$Si$_5$ follow the Curie-Weiss law (for temperatures above $\sim75$~K), $\chi({\rm T}) = \chi_0 + C/(T-\theta_{\rm p})$, where $\chi_0$ is a temperature-independent term, $C$ = $N_{\rm A}\mu^2_{\rm eff}/3k_{\rm B}$ is the Curie constant, and $\theta_{\rm p}$ is the Weiss temperature. By fitting the $1/[\chi(T)-\chi_0]$ curves above 75~K to the Curie-Weiss model, an effective magnetic moment of $\mu_{{\rm eff}} = 2.69 \mu_{\rm B}$/Ce, $\theta_{\rm p} = -95.6$~K, and $\chi_0 = 0.0016$ emu/mol$\cdot$Oe were obtained for $H\parallel c$, whereas $\mu_{{\rm eff}} = 2.61~\mu_{\rm B}$/Ce, $\theta_{\rm p} = -27.4$~K, and $\chi_0 = -0.0004$ emu/mol$\cdot$Oe were obtained for $H\perp c$. These values of the effective moments give a polycrystalline average of $2.64~\mu_{\rm B}$/Ce for Ce$_2$NiAl$_6$Si$_5$, close to $2.54~\mu_{\rm B}$/Ce expected for Ce$^{3+}$ free ions.
For $H \perp c$, the $\chi(T)$ of Ce$_2$NiAl$_6$Si$_5$ bends \red{downward} around 0.6~K, suggesting the onset of \red{AFM} order or the growth of dynamic \red{AFM} correlations.

The isothermal magnetization $M(H)$ of Ce$_2$NiAl$_6$Si$_5$ at 1.8~K is shown in Fig.~\ref{figure3}(b). The $M(H)$ is linear up to 7~T for $H \parallel c$, while for $H \perp c$, it is linear up to 1~T and then bends over at higher fields, without saturation up to 7~T. The polarized magnetic moment per Ce at 7~T is 0.44~$\mu_{\rm B}$ for $H \perp c$, which is much smaller than the free Ce$^{3+}$ ion moment. This may result from crystalline-electric field (CEF) effects, Kondo screening, or significant AFM interactions between the Ce moments. The $M(H)$ of Ce$_2$NiAl$_6$Si$_5$ for $H \perp c$ at various temperatures are displayed in Fig. \ref{figure3}(c). With increasing temperature, $M(\rm H)$ for $H \perp c$ becomes smaller in magnitude and the bending over is weakened.

\begin{figure}[htbp]
	\begin{center}
		\includegraphics[width=\columnwidth]{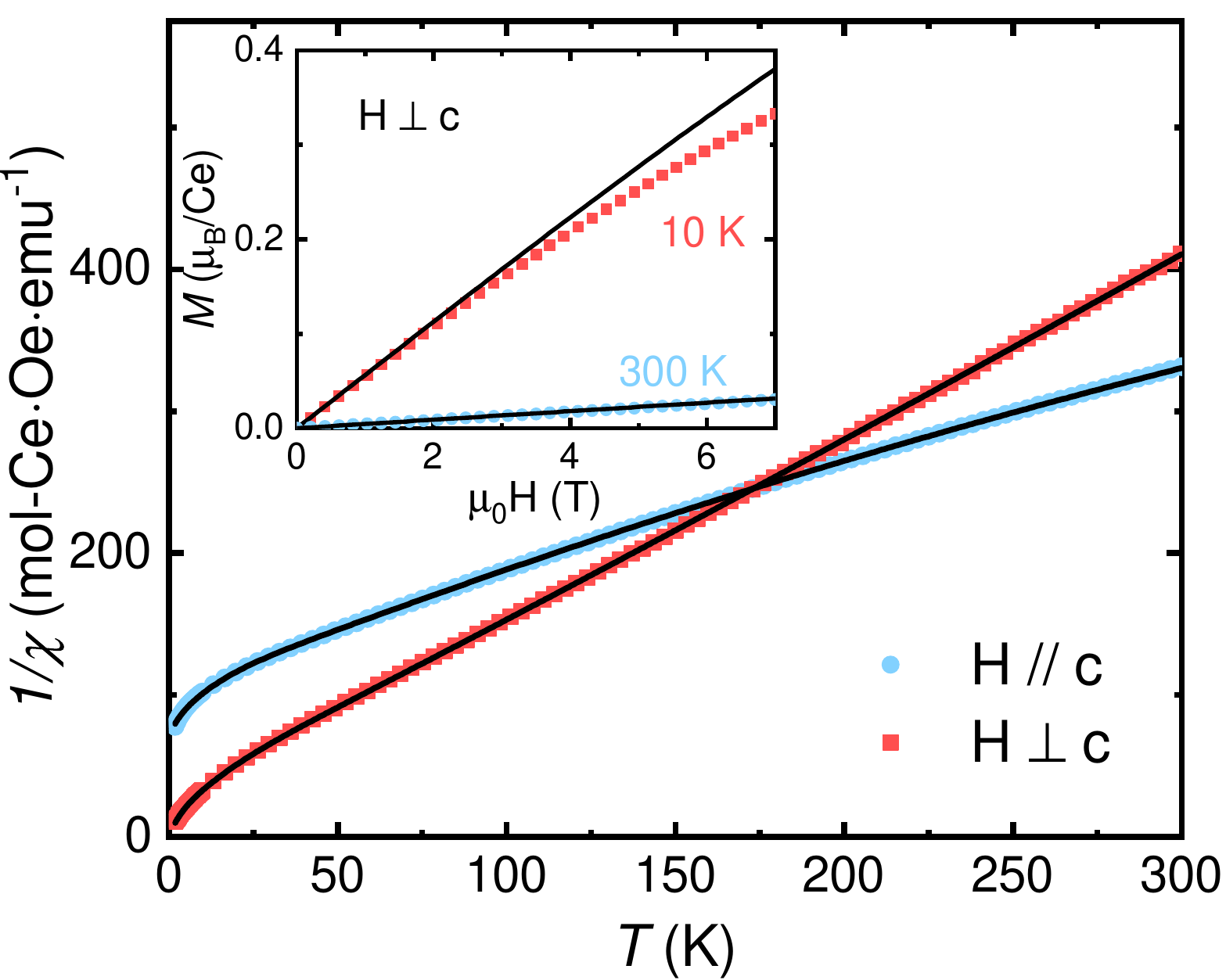}
	\end{center}
	\caption{(Color online)  $1/\chi(T)$ and $M(H)$ (inset) of Ce$_2$NiAl$_6$Si$_5$ fit to a crystalline-electric field (CEF) model (solid lines), as described in the text.}
	\label{figure4}
\end{figure}

By considering the local point group symmetry of the Ce ions, the anisotropic magnetic susceptibilities and the isothermal magnetization can be modeled with an orthorhombic CEF Hamiltonian:
\begin{equation}
    H_{\rm CEF} = B^0_2O^0_2 + B^2_2O^2_2 + B^0_4O^4_2 + B^2_4O^2_4 + B^4_4O^4_4
\end{equation}
where $B^n_i$ are the CEF parameters and $O^n_i$ are the Stevens operators \cite{KWHStevens1952}. $H_{\rm CEF}$ leads to a temperature-dependent magnetic susceptibility $\chi_{\rm CEF}$ of isolated Ce$^{3+}$ ions, and inclusion of interactions between the ions result in an effective susceptibility $\chi_{\rm eff} = \chi_{\rm CEF}/(1-\lambda\chi_{\rm CEF})+\chi_0$, where $\chi_0$ is an temperature-independent term and $\lambda$ parameterizes the effective mean-field interaction. The magnetic susceptibility incorporating the CEF Hamiltonian was calculated using Mantid Plot \cite{ArnoldO2014b}, and fit to the data by minimizing $X^2 = (\chi_{\rm calc} - \chi_{\rm obs})^2/\chi_{\rm calc}$ and allowing for anisotropic values of $\lambda$ and $\chi_0$, yielding $X^2 = 0.189$ for the optimal fit. The extracted parameters are $B^0_2 = 0.23$~meV, $B^2_2 = 0.69938$~meV, $B^0_4 = 0.00031$~meV, $B^2_4 = 0.04972$~meV, $B^4_4 = 0.07812$~meV, $\lambda_{\perp} = -1.17$~mol/emu, $\lambda_{\parallel} = -76.29$~mol/emu, $\chi^\perp_0 = -0.00008$~emu/mol, and $\chi^\parallel_0 = 0.00091$~emu/mol.
The energy levels and wave functions are summarized in Table~\ref{table2}, with the corresponding \red{fitted susceptibilities} shown in Fig.~\ref{figure4}. The CEF analysis shows that the ground state of Ce$_2$NiAl$_6$Si$_5$ is a doublet, with excited states at $\sim2.4$~meV and $\sim9.7$~meV. Further inelastic neutron scattering measurements will be desirable to confirm the CEF scheme of Ce$_2$NiAl$_6$Si$_5$.

\begin{table}[!ht]
	\renewcommand\arraystretch{1.4}
	\caption{Energy levels and wave functions of the CEF scheme of Ce$_2$NiAl$_6$Si$_5$.}
	\begin{tabular}{lllllll}
		\hline\hline
		$E$(meV)~ & $\ket{-5/2}$~ & $\ket{-3/2}$~ & $\ket{-1/2}$~ & $\ket{+1/2}$~ & $\ket{+3/2}$~ & $\ket{+5/2}$
		\\ \hline
		0 & -0.426 & 0.0 & 0.877 & 0 & -0.222 & 0.0 
		\\
		0 & 0.0 & 0.222 & 0.0 & -0.877 & 0.0 & 0.426
		\\
		2.42 & -0.467 & 0.0 & -0.003 & 0.0 & 0.884 & 0.0 
		\\
		2.42 & 0.0 & 0.884 & 0.0 & -0.003 & 0.0 & -0.467 
		\\
		9.73 & 0.775 & 0.0 & 0.481 & 0.0 & 0.411 & 0.0 
		\\
		9.73 & 0.0 & 0.411 & 0.0 & 0.481 & 0.0 & 0.775
		\\ \hline\hline
	\end{tabular}
	\label{table2}
\end{table}

\begin{figure}[htbp]
	\begin{center}
		\includegraphics[width=\columnwidth]{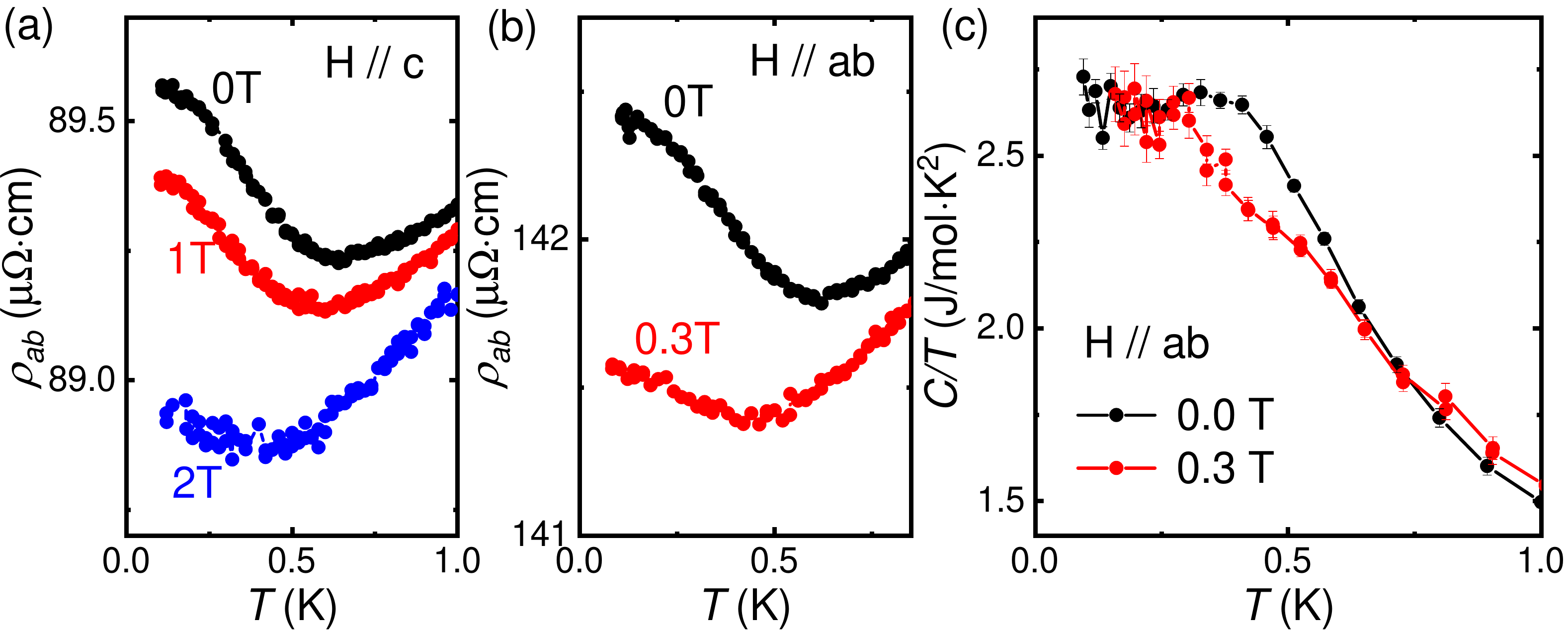}
	\end{center}
	\caption{(Color online)  \red{$\rho_{ab}$ of Ce$_2$NiAl$_6$Si$_5$ for fields (a) along the $c$-axis and (b) in the $ab$-plane. (c) $C(T)/T$ of Ce$_2$NiAl$_6$Si$_5$ for fields in the $ab$-plane.}}
	\label{figure5}
\end{figure}

\red{To confirm that the low-temperature anomalies in resistivity, magnetic susceptibility, and specific heat are associated with AFM ordering, the field-tuning of $\rho_{ab}$ and $C(T)/T$ are shown Fig.~\ref{figure5}. The $\rho_{ab}$ of Ce$_2$NiAl$_6$Si$_5$ for fields along the $c$-axis and in the $ab$-plane are displayed in Figs.~\ref{figure5}(a) and (b). While the resistivity upturn is suppressed towards lower temperatures in both cases, the response is highly anisotropic, with the suppression much faster for $ab$-plane fields, consistent with Ce$_2$NiAl$_6$Si$_5$ having an easy-plane magnetic anisotropy. Similarly, the bulge in heat capacity is also suppressed towards lower temperatures under applied field, as shown in Fig.~\ref{figure5}, consistent with the anomaly resulting from AFM ordering.}

Taken together, our characterizations of Ce$_2$NiAl$_6$Si$_5$ are consistent with a heavy fermion metal that orders below 0.6~K, although signatures of the \red{AFM} ordering are more subtle compared to well-ordered heavy fermion systems \cite{Fisk1991}. Compared to Ce$_2$NiAl$_6$Si$_5$, clearer signatures of magnetic ordering were observed in Ce$_2$NiAl$_7$Ge$_4$ at 0.8~K in specific heat. This suggests that Ce$_2$NiAl$_7$Ge$_4$ has a more robust magnetic order compared to Ce$_2$NiAl$_6$Si$_5$, although magnetic Bragg peaks were still too weak to be detected in neutron powder diffraction measurements \cite{PhysRevB.93.205141}. Therefore, both Ce$_2$NiAl$_6$Si$_5$ and Ce$_2$NiAl$_7$Ge$_4$ are likely in proximity to \red{AFM} QCPs, with Ce$_2$NiAl$_6$Si$_5$ having weaker magnetic order and closer to the QCP. Further neutron single crystal diffraction measurements are desired to elucidate the magnetic properties of Ce$_2$NiAl$_6$Si$_5$.

In summary, we successfully grew single crystals of Ce$_2$NiAl$_6$Si$_5$ using the self-flux method, which crystallize in a centrosymmetric tetragonal structure with space group $P4/nmm$ (No.~129). Electrical resistivity and magnetic susceptibility measurements suggest the onset of \red{AFM} ordering or dynamic \red{AFM} correlations around 0.6~K, although the corresponding anomaly in specific heat appears at a slightly lower temperature of 0.4~K.
The large low temperature $C(T)/T$ and reduced entropy indicates a significant Kondo effect. These results indicate that Ce$_2$NiAl$_6$Si$_5$ is a \red{quasi-2D} layered heavy fermion metal close to quantum criticality, and motivate further systematic studies of its tunability under pressure or magnetic field. 

\section{acknowledgments}
The work at Zhejiang University was supported by the Pioneer and Leading Goose R$\&$D Program of Zhejiang (2022SDXHDX0005), the National Key R$\&$D Program of China (Grants No. 2022YFA1402200, No. 2023YFA1406100), the Key R$\&$D Program of Zhejiang Province, China (Grant No.2021C01002), the National Natural Science Foundation of China (Grants No. 12034017, No. 12274363, No.12350710785), the Fundamental Research Funds for the Central Universities (Grant No. 226-2024-00068).

\bibliography{ref}

\end{document}